\documentclass[12pt]{iopart}
\usepackage{iopams}

\usepackage{cite}
\usepackage{graphicx}
\usepackage{subfigure}
\usepackage{makecell}
\usepackage{threeparttable}
\usepackage{siunitx}
\usepackage{hyperref}
\usepackage{xcolor}
\definecolor{bluee}{rgb}{0.062,0.086,0.796}
\hypersetup{colorlinks=true,linkcolor=bluee,urlcolor=bluee,anchorcolor=bluee,citecolor=bluee,CJKbookmarks=True}

\begin{document}
\title[Picometer-level quadrangle optical bonding bench for testing interferometric technologies in TianQin]{Picometer-level quadrangle optical bonding bench for testing interferometric technologies in TianQin}
\author{Hao Yan*, Xiang Lin}
\address{Center for gravitational experiment, MOE Key Laboratory of Fundamental Physical Quantities Measurements, The School of Physics, Huazhong University of Science and Technology, Wuhan 430074, China}
\author{Siyuan Xie}
\address{MOE Key Laboratory of TianQin Mission, TianQin Research Center for Gravitational Physics \& School
of Physics and Astronomy, Frontiers Science Center for TianQin, Gravitational Wave Research Center of CNSA, Sun Yat-sen University (Zhuhai Campus), Zhuhai 519082, China}
\ead{yanhao2022@hust.edu.cn}
\begin{abstract}
Interferometric techniques are crucial for space-based gravitational wave detection, requiring a picometer-level stable optical bench, precise phasemeter, interstellar transponder low-light phase locking, and laser sideband communication. These technologies must be rigorously tested on the ground before deployment in space. The AEI group has previously developed a picometer-stable hexapod optical bench to verify the linearity and precision of phase extraction for LISA. In this paper, we introduce a quadrangle quasi-monolithic optical bench aimed at simplifying the system and expanding the range of tested interferometric techniques for TianQin. Experimental results demonstrate that the system achieves picometer-level optical pathlength stability and phase resolution over a large dynamic range. In the laser transponder link test, the light phase-locked residual noise is lower than ${\rm 10^{-4}\,rad/Hz^{1/2}}$ above millihertz frequency range, and the laser sideband modulation has no significant coupling to the measurements in the ${\rm mHz-Hz}$ band. These results provide critical technical validation for the implementation of future gravitational wave detection in space. 
\end{abstract}

\submitto{\CQG}

\section{Introduction}
The first successful detection of gravitational waves by the ground-based Laser Interferometer Gravitational-Wave Observatory (LIGO)\cite{LIGO} opens a new era in gravitational wave astronomy. Long-baseline space-based gravitational wave detectors, such as LISA\cite{LISA}, TianQin\cite{TianQin}, and Taiji\cite{Taiji}, are designed to address the limitations of ground-based observatories by offering enhanced sensitivity in the low-frequency ${\rm mHz-Hz}$ band. Given the extremely weak nature of gravitational wave signals, detectors must achieve extraordinarily high displacement sensitivity. Unlike ground-based detectors like LIGO\cite{LIGO}, VIRGO\cite{LIGO2}, and KAGRA\cite{KAGRA}, which utilize homodyne interferometry (based on DC light intensity detection), space-based gravitational wave detectors employ heterodyne interferometry, which involves beat-frequency phase measurements\cite{LISA}. This approach allows for the detection of tiny changes in distance between two free-floating test masses at long baselines, with sensitivity reaching the picometer level. Achieving such precision requires the use of ultra-stable optical benches, high-precision phasemeters, interstellar transponder low-light phase locking, and laser sideband communication.

The optical bench (OB) is one of the most crucial components of the Space Gravitational Wave Detector. It is responsible for establishing the link between the two free-floating test masses (TM) across satellites, forming the TM-OB-OB-TM link \cite{bench}. For the TianQin mission, maintaining the light-path stability of the optical benches below  ${\rm 1\,pm/Hz^{1/2}}$ is essential \cite{TianQin}. Achieving this goal requires significant effort. First, low-expansion materials, such as ULE (Ultra-Low Expansion) glass, are utilized for the substrate of the optical bench to minimize thermal coupling noise. Second, optical components, including collimators, mirrors, beam-splitters, and waveplates, are bonded to the optical bench to create a quasi-monolithic structure, ensuring both thermal and mechanical stability \cite{bench2}.

In the space-based gravitational wave detection, the phasemeter is responsible for measuring all heterodyne interference signals with exceptional precision \cite{phasemeter1,phasemeter2,phasemeter3}. The phasemeter must achieve an accuracy of ${\rm 1\,\mu rad/Hz^{1/2}}$ down to the millihertz frequency range. Additionally, the phasemeter must handle the Doppler effect resulting from the relative motion between satellites, which causes shifts in the received signal frequencies. Consequently, the phasemeter needs to cover a broad dynamic heterodyne frequency range from ${\rm 5-25\, MHz}$ while maintaining high linearity. This ensures accurate measurement of the data amidst the frequency variations introduced by satellite motion.

For long-baseline space gravitational wave detection, inter-satellite transponder laser links are utilized through weak light phase-locking techniques. In this setup, a master satellite transmits a high-power laser at the watt level. The slave satellites then use weak-light phase-locking to synchronize their local lasers with the incoming laser signal, which is received at the sub-nanowatt level. After phase-locking to this weak signal, the slave satellites transmit a high-power laser back to the master satellite\cite{laser-link}. The return laser then interferes with the local laser of the master satellite. Ultimately, the inter-satellite distance change is obtained from two phase signals.

In addition, space-based gravitational wave detection requires precise laser ranging and clock synchronization \cite{PRN,clk}, particularly during the data post-processing phase of time-delayed interferometry (TDI)\cite{Tinto2020}. To meet these demands, the technique involves phase modulating the laser with pseudo-random (PRN) codes and clock sidebands. This modulation technique is crucial as it enables interstellar laser sideband communication, ensuring accurate laser ranging, data transmission, and clock synchronization between the satellites. 

Ground-based experimental validation is critical for the development of space-borne gravitational wave detectors, as the optical bench and phasemeter must be tested together under controlled conditions. However, one major challenge is ensuring that the system's stability and sensitivity are maintained during testing. Several methods are used to validate these systems, such as employing high-precision commercial phasemeters with narrow bandwidths, conducting differential phase measurements of identical interferometric signals through balanced detection or beam splitting, or using an electronic signal generator to test the phasemeter\cite{phasemeter1,phasemeter2,phasemeter3}. One of the most effective approaches is to construct an ultra-stable optical bench. Instead of relying on free-moving test masses, as would be used in space, a fixed reflector can be implemented to verify the optical bench's stability on the ground.

The LISA project has made significant strides in developing such stable optical benches. In 2003, D. A. Shaddock, B.C. Young and A. Abramovici from NASA's Jet Propulsion Laboratory (JPL) developed a rigid quadrangle interferometer bench made entirely of ultra-low expansion (ULE) glass through optical contacting. Preliminary results indicate optical path length fluctuations of approximately ${\rm 100\,pm/Hz^{1/2}}$ for frequencies between ${\rm 1\,mHz}$ and ${\rm 1\,Hz}$ \cite{Daniel2003}. This work laid the groundwork for future developments in optical bench stability, particularly for space-based detectors like LISA.

In 2006, T. S. Schwarze, G. F. Barranco, G. Heinzel, etc., from AEI made significant progress by introducing a hexagonal quasi-monolithic optical bench \cite{Hexagonal}. This bench was designed to perform a three-signal test, enabling the verification of interferometric techniques, particularly the phase extraction linearity and precision. Their experimental results demonstrated that the hexagonal bench was stable down to picometer levels and allowed the use of a free-running master laser. This optical testbed was the first to achieve ${\rm MHz}$ signal extraction with ${\rm \mu cycle/Hz^{1/2}}$ precision, serving as a key validation tool for LISA's metrology chain. In 2022, the AEI group further achieved experimental verification of the inter-satellite clock synchronization scheme and absolute ranging down to LISA performance levels using the same hexagonal optical bench \cite{Yamamoto1,Yamamoto2}.

Inspired by LISA’s optical bench testbed designs and based on the requirements of the TianQin project\cite{Cao24}, we developed a quadrangle quasi-monolithic optical bench. Our simplified design maintains the necessary stability while expanding the range of interferometric techniques that can be tested, including picometer-level optical pathlength stability, phasemeter, interstellar transponder low-light phase locking, and sideband communication. This paper details the design, experimental setup, and testing of our optical bench, which serves as a critical validation platform for the future TianQin space gravitational wave detection mission. The paper is organized as follows: Section II outlines the theoretical modeling of the optical design, Section III presents the experimental setup and test results, and Section IV offers a summary and future directions.

\section{Optical design}
\begin{figure}[ht!]
\centering\includegraphics[width=0.8\textwidth]{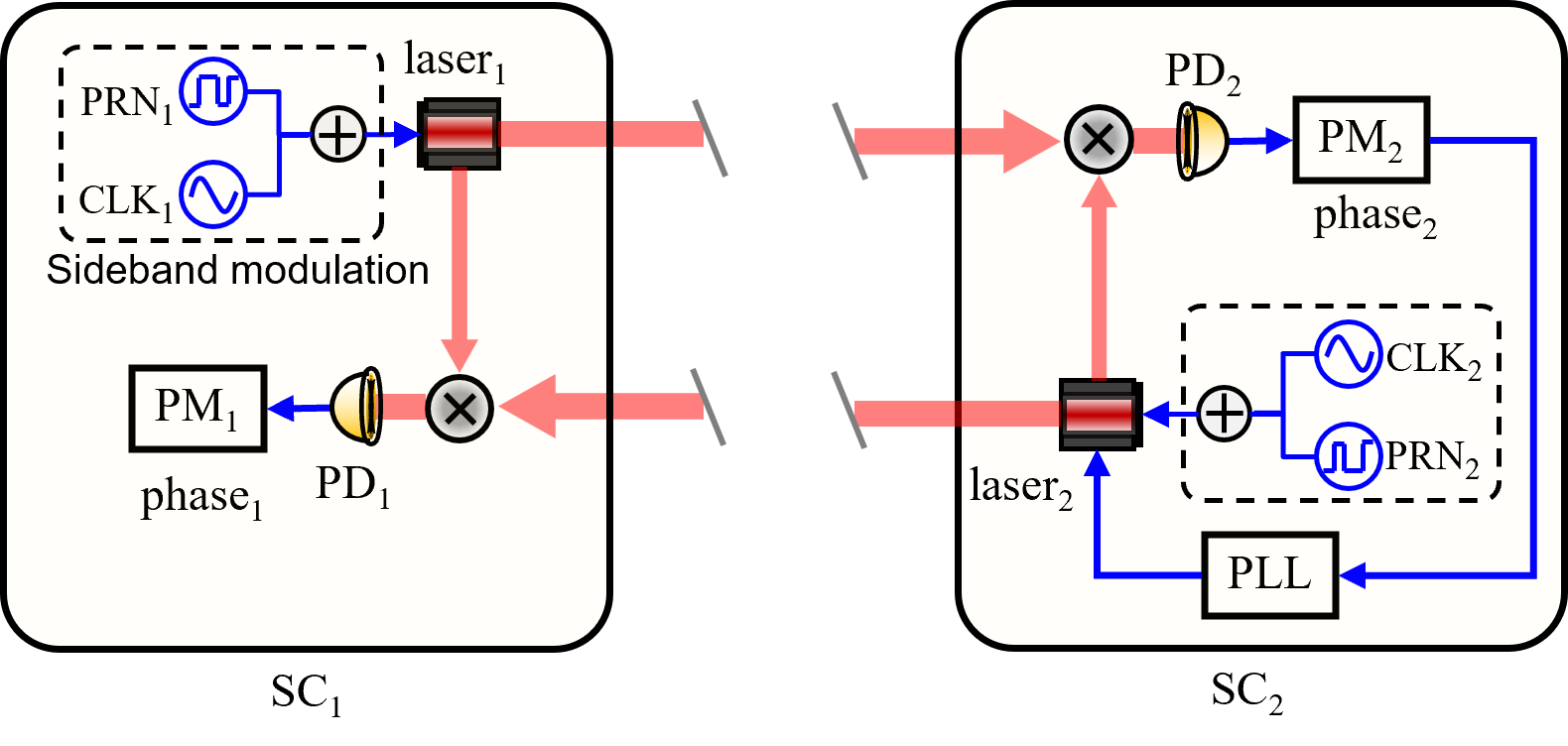}
\caption{\label{fig1} Schematic of interstellar transponder laser interferometric link for space gravitational wave detection.}
\end{figure}

For long-baseline space gravitational wave detection, inter-satellite transponder laser links are utilized \cite{PRN,clk}, as depicted in figure \ref{fig1}. The master satellite ${\rm SC_1}$ transmits a high-power laser, typically at the watt level. The slave satellite ${\rm SC_2}$ uses an interferometer and a phasemeter (PM) to measure the receiving phase signal. The phase-locking loop (PLL) is employed to synchronize its local laser (${\rm laser_2}$) with the incoming weak-light signal, which is received at sub-nanowatt power levels. Once phase-locking is achieved, the slave satellite transmits a high-power laser back to the master satellite. The phase signal (${\rm phase_1}$) carries information about interstellar distance variations. Additionally, for laser sideband communication, phase modulation is applied to both the master and slave satellite lasers. This includes pseudo-random code (PRN) modulation for absolute inter-satellite laser ranging and data communication, as well as clock sideband modulation for clock synchronization.

\subsection{bench lightpath stability and phasemeter}
\begin{figure}[ht!]
\centering\includegraphics[width=1\textwidth]{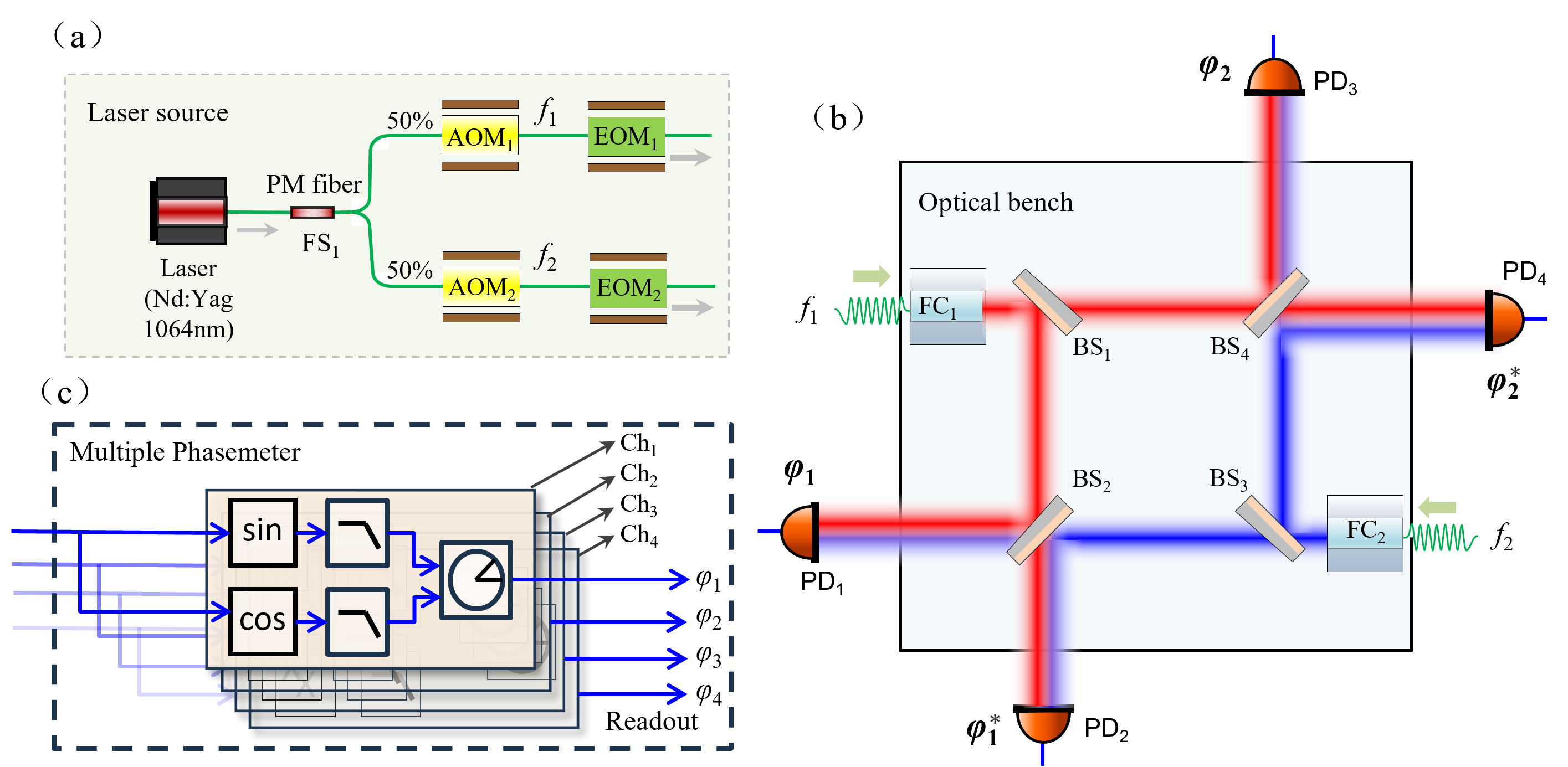}
\caption{\label{fig2} Schematic of the optical two-signal setup: panel (a) shows the laser sources. A laser of wavelength ${\rm 1064\,nm}$ is $50\%:50\%$ divided into two beams via a fibre splitter. Each beam passes through an acousto-optic modulator (AOM) and an electro-optic phase modulator (EOM), respectively. All fiber components are polarization maintaining (PM). The final output is two equal power laser beams with frequencies $f_1$ and $f_2$, respectively. Panel (b) shows quadrangle quasimonolithic optcial bench. Two heterodyne laser beams $f_1$ and $f_2$ are injected by fiber collimators, ${\rm FC_1}$ and ${\rm FC_2}$. The two heterodyne lasers are beam-split on ${\rm BS_1}$ and ${\rm BS_3}$, and beam-combined interfering on ${\rm BS_2}$ and ${\rm BS_4}$. The interference beat signals are detected by four photodetectors (${\rm PD_1}$, ${\rm PD_2}$, ${\rm PD_3}$, and ${\rm PD_4}$) and converted into an electrical signal. Panel (c) illustrates the multichannel phasemeter which is responsible for extracting the phases of the four interfering signals, i.e., $\varphi_1$, $\varphi_2$, $\varphi_3$, and $\varphi_4$.}
\end{figure}

An optical two-signal scheme is designed to test the lightpath stability and the phasemeter. figure \ref{fig2} depicts the conceptual design of the optical setup. The laser sources are frequency-shifted by two acousto-optic modulators (AOMs) and then phase-modulated by two electro-optic phase modulators (EOMs) as shown in figure \ref{fig1}(a). The heterodyne laser beams are injected into the quadrangle optical bench, figure \ref{fig1}(b),by polarization maintaining fiber. Four $50\%:50\%$ beamsplitters (${\rm BS_1}$, ${\rm BS_2}$, ${\rm BS_3}$, and ${\rm BS_4}$) are distributed in a centripetal square on the optical bench. The two interfering signals were precisely designed with equal arm lengths in order to common-mode suppress coupling noise (laser frequency fluctuations, ambient temperature perturbations, etc.). A balanced detection design is employed. The four interference signals are detected by a multiple channels phasemeter shown in figure \ref{fig1}(c). 
 
Assuming that $\{\varphi_1$,\,$\varphi_1^\ast\}$ and $\{\varphi_2$\,$\varphi_2^\ast\}$ are the four phase signals of the two balance detections measured by the four photodetectors $\{{\rm PD_1}$, ${\rm PD_2}$, ${\rm PD_3}$, ${\rm PD_4}\}$ shown in figure \ref{fig1}(b):
\begin{eqnarray}
    \varphi_1=\varphi_{f_1}-\varphi_{f_2}+\varphi_{path_1}, \label{eq1}\\
    \varphi_1^\ast=\varphi_{f_1}-\varphi_{f_2}+\varphi_{path_1}+\pi, \label{eq2}\\
    \varphi_2=\varphi_{f_2}-\varphi_{f_1}+\varphi_{path_2},\label{eq3}\\
    \varphi_2^\ast=\varphi_{f_2}-\varphi_{f_1}+\varphi_{path_2}+\pi,\label{eq4}    
\end{eqnarray}
where $\{\varphi_{f_1},\ \varphi_{f_2}\}$ are inertial phases of the two heterodyne laser beams from fiber collimators ${\rm FC_1}$ and ${\rm FC_2}$, respectively. $\{\varphi_{path_1},\ \varphi_{path_2}\}$ are unequal-arm lightpaths of the two heterodyne interference signals and satisfy:
\begin{eqnarray}
    \varphi_{path_1}\approx\varphi_{path_2}\approx 0. \label{eq5}
\end{eqnarray}
The additional $\pi$ is caused by the balanced detection. 

The above measured phases can be combined to form a zero output by:
\begin{eqnarray}
    \varphi_0=\varphi_1+\varphi_2=\varphi_1^\ast+\varphi_2^\ast, \label{eq6}
\end{eqnarray}
or a $\pi$ output from balance detection by:
\begin{eqnarray}
    \varphi_{\pi}=\varphi_1^\ast-\varphi_1=\varphi_2^\ast-\varphi_2, \label{eq7}
\end{eqnarray}
which are the main measurements and in which the initial phases ideally cancel. The $\varphi_0$ and $\varphi_{\pi}$ can be used for lightpath stablity and phasemeter testing, respectively. 

\subsection{Laser transponder phase locking and sideband communication}
\begin{figure}[ht!]
\centering\includegraphics[width=1\textwidth]{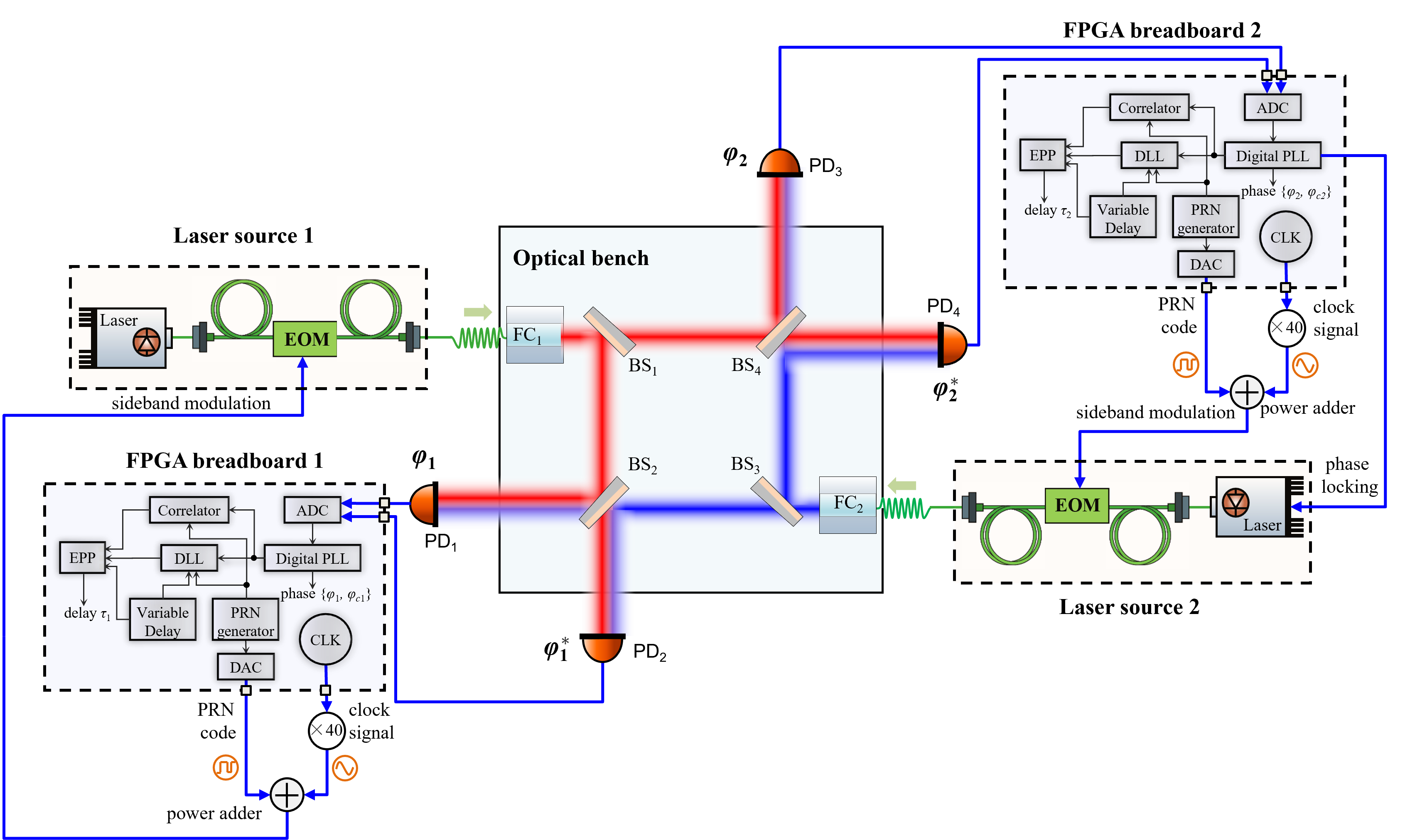}
\caption{\label{fig3} Schematic of the laser transponder phase locking and sideband communication test. Two identical sets of laser sources and FPGA breadboards are employed to simulate two satellites. The two lasers are first sideband phase modulated by independent EOMs (${\rm EOM_1}$ and ${\rm EOM_2}$) and then injected into the interferometer platform by collimators (${\rm FC_1}$ and ${\rm FC_2}$). The FPGA breadboards are responsible for laser sideband modulation and phase locking in addition to interferometric phase measurements.}
\end{figure}

To further simulate the actual in-orbit satellites for the detection of gravitational waves in space, laser transponder phase-locking and sideband communication techniques need to be tested. We adapted the optical system design based on the interstellar schematic of figure \ref{fig1}. The new design is shown in figure \ref{fig3}. Two identical sets of laser sources and field-programmable gate arrays (FPGA) breadboards are employed to simulate two satellites. Laser sources 1 and 2 represent the master and slave lasers, respectively. The slave laser is kept heterodyne phase locked to the master laser by the phase signal $\varphi_2$, and the phase-locked residuals are measured by the phase signal $\varphi_1$.

\begin{figure}[ht!]
\centering\includegraphics[width=0.85\textwidth]{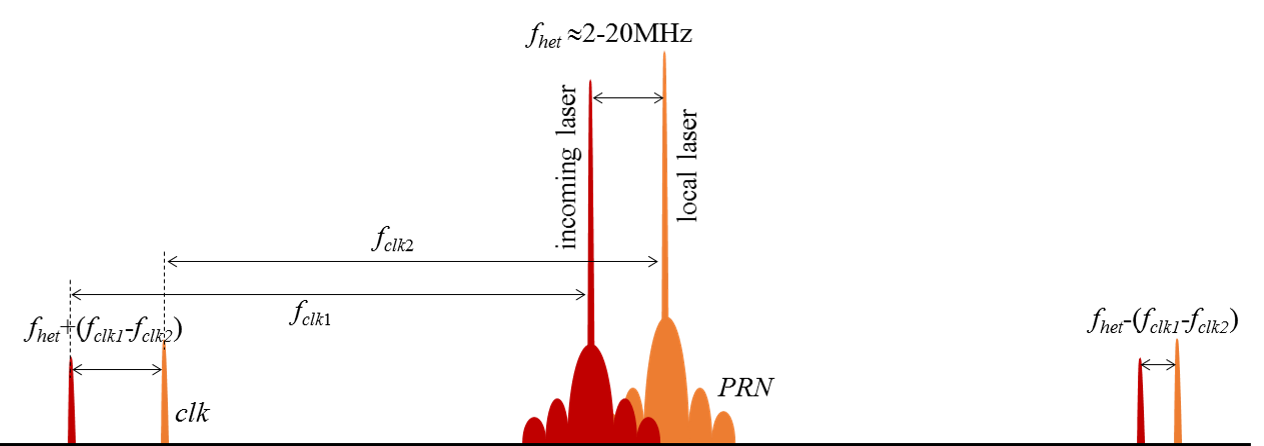}
\caption{\label{fig4} Spectral structure of interstellar laser interference signals after laser sideband modulation.}
\end{figure}

Laser sideband communication is achieved by EOM phase modulation as on-orbit. The carrier laser light field carrying a sideband modulation can be expressed as:
\begin{eqnarray}
    E(t)=E_0\cdot exp\left[ i\omega_0t+i\varphi_0+i\varphi_{clk}(t)+i\varphi_{prn}(t)\right], \label{eq8}
\end{eqnarray}
where the $\omega_0$ and $\varphi_0$ are the frequency and inertial phase of the carrier laser, respectively. $\varphi_{clk}(t)$ and $\varphi_{prn}(t)$ are the clock and PRN phase-sideband modulation, respectively, and are satisfied:
\begin{eqnarray}
    \varphi_{clk}(t)=m_{clk}cos(\omega_{clk}t), \label{eq9}\\
    \varphi_{prn}(t)=m_{prn}\sum_{n=-\infty}^{+\infty }c_{prn}p(t-nT_{prn}), \label{eq10}   
\end{eqnarray}
where $m_{clk}$ and $\omega_{clk}$ are the depth and frequency of clock phase modulation, respectively. $m_{prn}$ and $p(t)$ are the depth and pulse sequence with period $T_{prn}$ of PRN phase modulation, respectively. $c_{prn}$ is a binary phase-shift keyed PRN code sequence. figure \ref{fig4}
The impact of modulation on the detection of gravitational waves in space can be initially assessed by comparing the phase-locked residuals before and after laser sideband modulation.

\section{Experiment}
\subsection{Prototype setup}
Experiments were conducted using the schematics of optical design depicted in figure \ref{fig2} and figure \ref{fig3}. To achieve high stability and precision, we have constructed an all-glass quasi-monolithic compact quadrangle optical bench using UV-adhesive bonding\cite{Lin2023},  and tested the bench in a temperature-controlled chamber with fluctuations $\delta T{\rm <1\,mK/Hz^{1/2}}$ at ${\rm 1\,mHz}$. In order to suppress the coupling noise, the laser frequency and laser power are controlled at levels of $\delta f/f{\rm<10^{-12}\,/Hz^{1/2}}$ and $\delta P/P{\rm<10^{-3}\,/Hz^{1/2}}$ at ${\rm 1\,mHz}$, respectively, via PID closed-loop feedback. 

\begin{figure}[ht!]
\centering\includegraphics[width=0.6\textwidth]{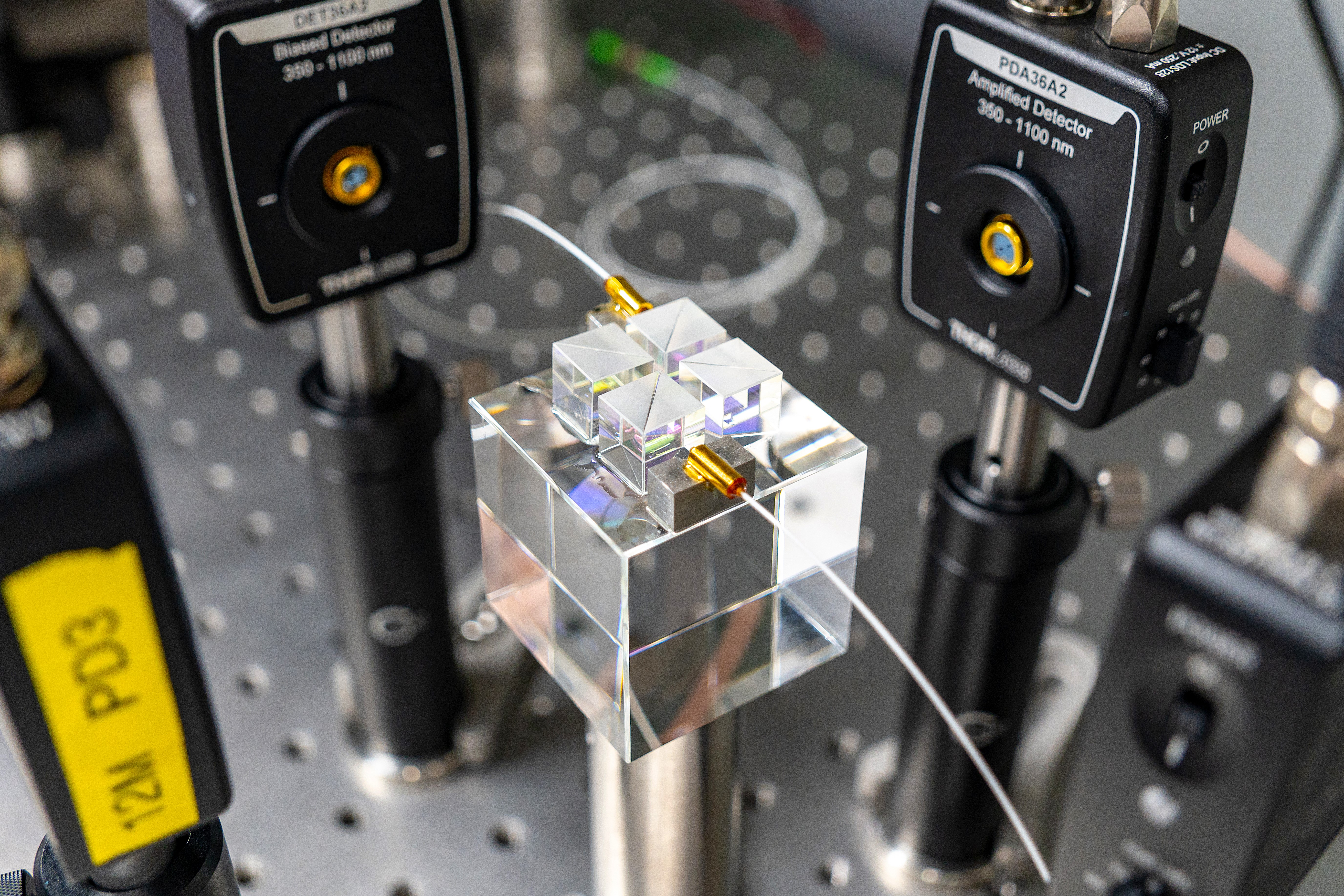}
\caption{\label{fig5} Photograph of the experimental prototype setup. Two injectors and four beamsplitters are bonded to the optical bench by UV adhesive. The four interfering signals were measured by four identical photodetectors from Throlabs.}
\end{figure}

\begin{table}
\caption{\label{tab1}The key parameters of the experimental setup.}
\centering
\begin{tabular}{cccccccc}
\hline
 Parameters&Value\\
\hline
laser wavelength& $1064\,\mathrm{nm}$\\
beam waist diameter& $\sim 1\,\mathrm{mm}$\\
input beam power& $\sim 1\,\mathrm{mW}$\\
size of the quadrangle bench (baseplate) & ${\rm 40\,mm\times 40\,mm\times 40\,mm}$\\
material of the optical bench and components & fused silica\\
optical bonding method & UV adhesive gluing\\
PRN sideband modulation depth & 0.1\,rad\\
PRN code bit rate & 1.5\,MHz\\
clock sideband modulation depth & 0.46\,rad\\
clock sideband frequency& 1\,GHz\\
\hline
\end{tabular}
\end{table}

The photograph of the experimental prototype setup is shown in figure \ref{fig5}. The quadrangle optical bonding bench, including the baseplate and optical components (fiber collimators and beamsplitters), are made of fused silica with dimensions of ${\rm 40\,mm\times 40\,mm\times 40\,mm}$. The optical bench is mounted on the top of a two-dimensional adjustment platform for testing. A Nd:YAG laser operating at a wavelength of ${\rm 1064\,nm}$ with a power of ${\rm 1\,mW}$ is connected to a fiber splitter. The phasemeter undertest is based on high-precision digital FPGAs implementing digital phase-locked loops (DPLLs). The heterodyne beat-note signals are digitalized by parallel analog-to-digital converter(ADC) channels (${\rm 100\,MHz}$), and are phase tracked by independent  numeric controlled oscillators (NCOs), which are realized by DPLLs\cite{yan2022all}. The key parameters of the experimental setup are provided in Table \ref{tab1}. 

\subsection{Test results}
\begin{figure}[ht!]
\centering\includegraphics[width=0.6\textwidth]{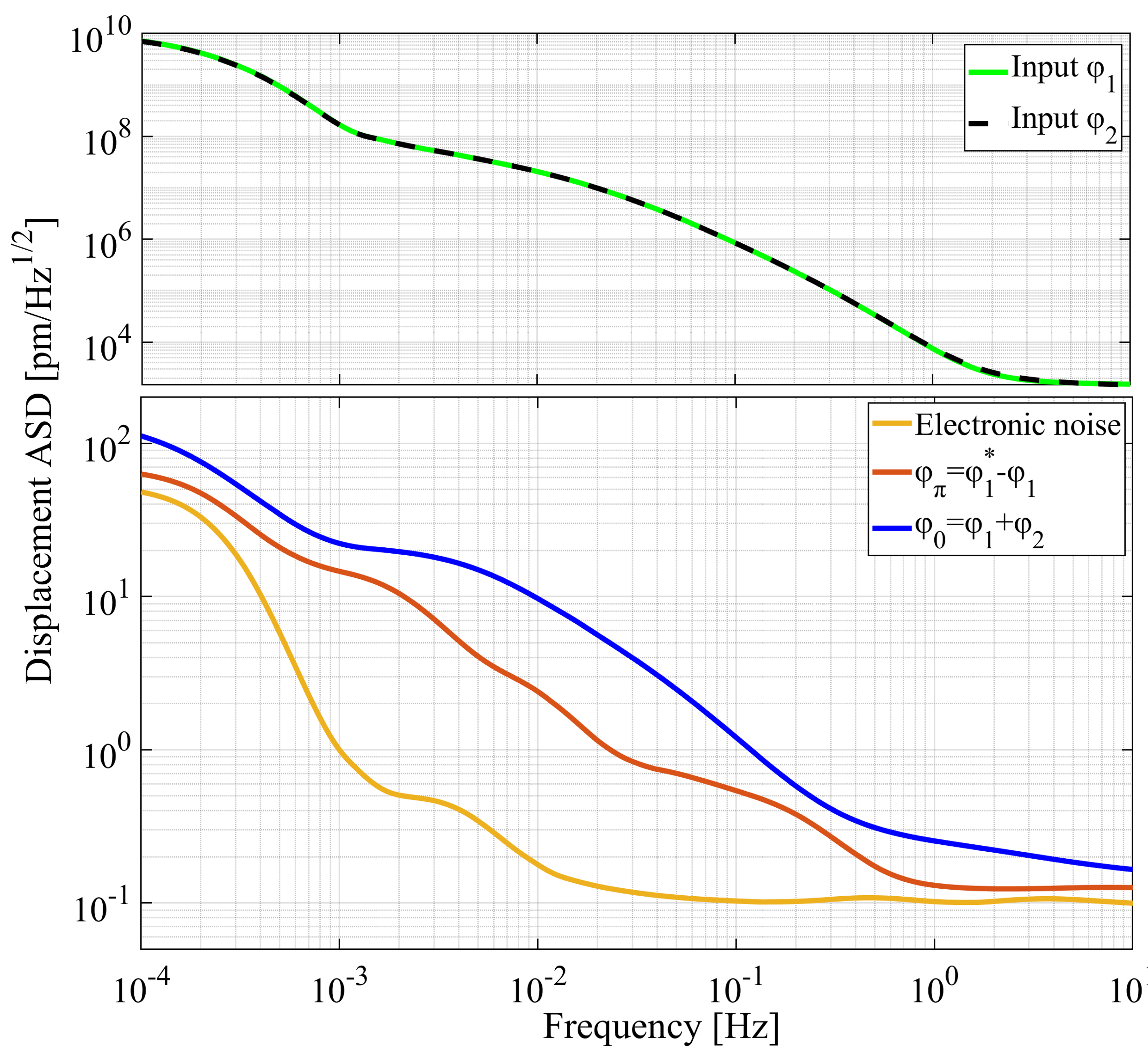}
\caption{\label{fig6} Test results of bench lightpath stability and phasemeter sensitivity with injection phase noise.  A dynamic range of 4–8 orders of magnitude can be computed from the top input signals ($\varphi_1$ and $\varphi_2$) and the two-signal combinations, $\varphi_0$ (blue) and $\varphi_\pi$ (red). The solid yellow line shows the electronic noise floor of the FPGA phasemeter.}
\end{figure}

The testing results of bench lightpath stability and phasemeter sensitivity are shown in figure \ref{fig6}. A picometer-level optical pathlength stability of the optical bench itself and phase resolution over a large dynamic range are acieved. The single input phases (top black and green lines) were generated to resemble a space orbit-like signal shape via the AOM frequency modulation (figure \ref{fig2}), with a phase noise of ${\rm 0.1\,rad/Hz^{1/2}}$ at ${\rm 1\,Hz}$ and a $1/f$ behavior dominating below ${\rm 10\,Hz}$. This corresponds to a dynamic range of 4–8 orders of magnitude from the top input signals ($\varphi_1$ and $\varphi_2$) and the two-signal combinations, $\varphi_0$ and $\varphi_\pi$.

The light pathlength stability of the quadrangle quasi-monolithic optical bench reaches a level of ${\rm 30\,pm/Hz^{1/2}}$ at ${\rm 1\,mHz}$ and ${\rm 0.3\,pm/Hz^{1/2}}$ at ${\rm 1\,Hz}$, as shown in the solid blue line (two-signal combinations, $\varphi_0$). It is a factor of ${\rm 3-5}$ higher than the balance detection combinations $\varphi_\pi$ (solid red curve) over the full frequency band, which measures the optical noise floor of the FPGA phasemeter. The yellow curve shows the purely electronic noise floor, i.e. the signal generator directly connected to the phasemeter. It can be concluded that the light pathlength stability of the bench is still the main noise source, and photoelectric detection noise is the main limit for the phasemeter in the ${\rm mHz-Hz}$ band. 

\begin{figure}[ht!]
\centering\includegraphics[width=0.7\textwidth]{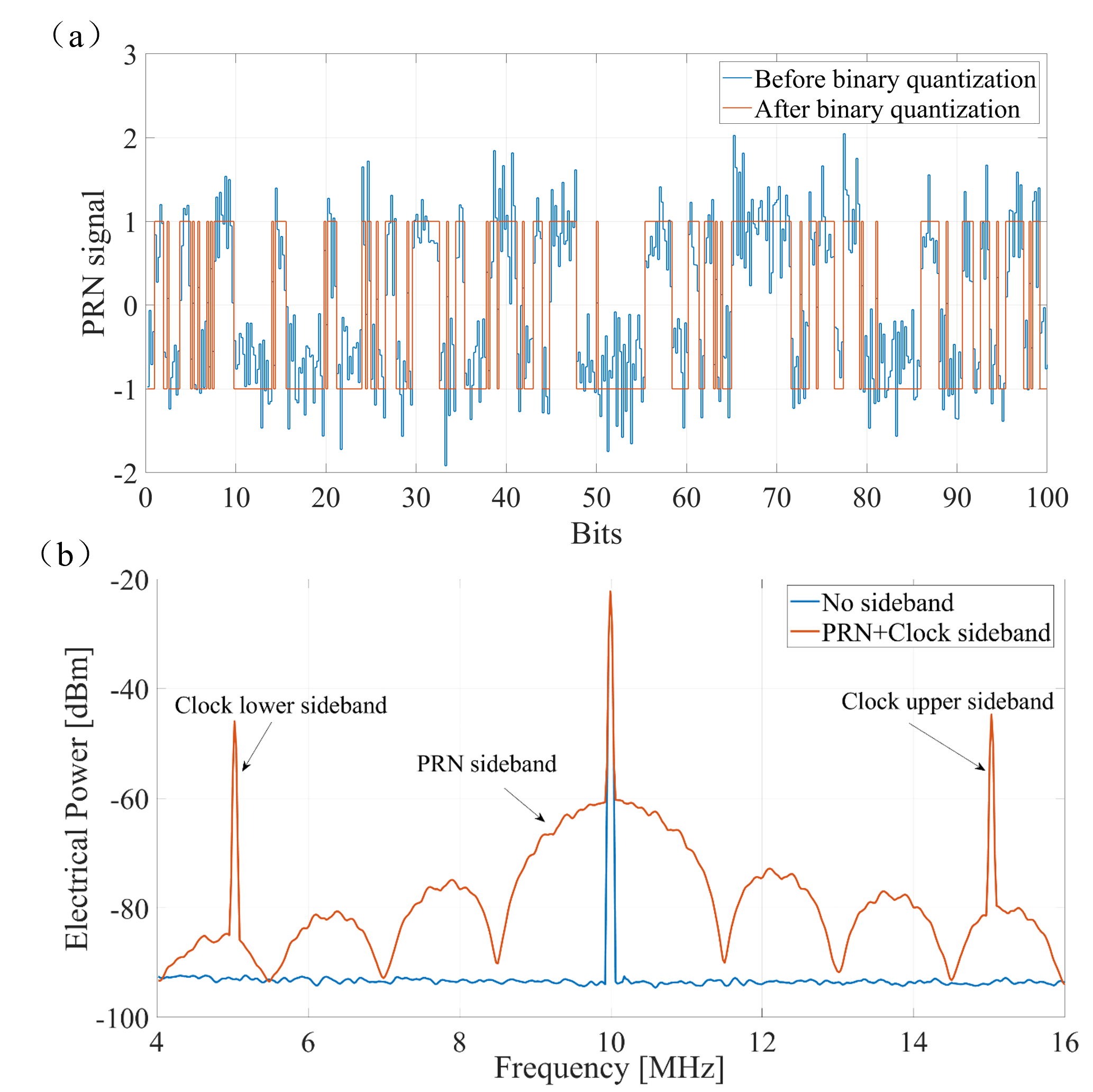}
\caption{\label{fig7} PRN code and Clock sideband signals. (a) PRN signal before and after binary quantization, (b) power spectrum distribution of the beat note signal.}
\end{figure}

\begin{figure}[ht!]
\centering\includegraphics[width=0.7\textwidth]{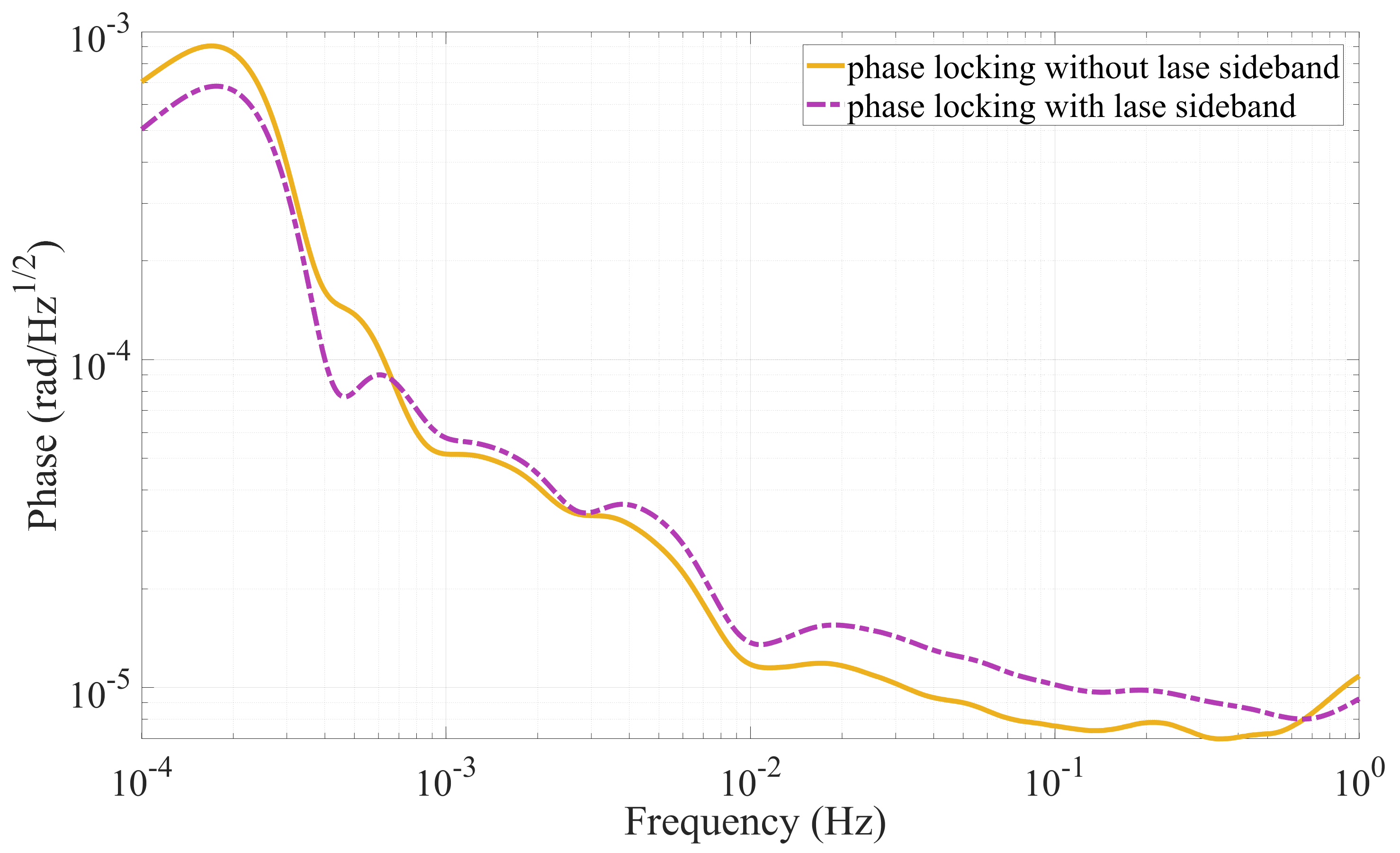}
\caption{\label{fig8} Test results of laser transponder phase-locking and sideband communication. The solid yellow line and dotted purple line show the phase-locking residuals before and after laser sideband modulation (PRN sideband and clock sideband).}
\end{figure}

The laser transponder phase-locking and sideband modulation were evaluated, as shown in Figure \ref{fig3}. Two laser sources and FPGA breadboards were employed for this test. Laser source 2 was phase-locked to laser source 1 via the heterodyne interference signal, $\varphi_2$, while the phase-locking residuals were measured using $\varphi_1$. The PRN signal and power spectrum of the beat note, including the laser sideband (PRN, clock sideband), are illustrated in Figure \ref{fig7}. As seen in Figure \ref{fig7}(a), the PRN code signal must be quantized into a binary format to enable high-speed operations. Although a certain level of bit error rate is present, the PRN signal effectively maintains delay lock and achieves accurate absolute ranging, as discussed in detail in \cite{PRN}. Figure \ref{fig7}(b) displays the power spectrum of the beat note with laser sideband modulation. To clearly distinguish between multiple signals, the carrier beat frequency was set to ${\rm 10,MHz}$, with the two clock sideband beat frequencies (representing the beat note of the two inter-satellite clock sidebands) set to ${\rm 5,MHz}$ and ${\rm 15,MHz}$, respectively.

The phase-locking test results are presented in Figure \ref{fig8}. The solid yellow line represents the phase-locking residuals without sideband modulation, demonstrating a phase noise of less than ${\rm 10^{-4},rad/Hz^{1/2}}$ above the ${\rm 1,mHz}$ frequency band. The dotted purple line shows the phase-locking residuals with the inclusion of laser phase sideband modulation, specifically the PRN code sideband and clock sideband modulation, as illustrated in Figure \ref{fig4}. A comparison of the two tests suggests that laser sideband modulation does not significantly couple into the measurements within the ${\rm mHz-Hz}$ band.

\section{Discussion and conclusion}
In this paper, we developed a quadrangle quasi-monolithic optical bench designed to test various interferometric techniques critical for space-based gravitational wave detection, particularly the TianQin mission. These techniques include picometer-level optical pathlength stability, high-precision phasemeter performance, interstellar transponder low-light phase locking, and laser sideband communication. Two experimental setups were constructed using the same quadrangle bench to test different aspects of these techniques. The first setup was dedicated to evaluating optical pathlength stability and phasemeter precision, utilizing a shared laser source, akin to that found within a single satellite. Heterodyne frequency modulation was achieved via a pair of acousto-optic modulators (AOMs). Experimental results confirmed that the optical bench maintained picometer-level pathlength stability, and the phase resolution was sustained across a large dynamic range. The second setup was used to assess interstellar transponder phase-locking under low-light conditions and laser sideband communication. Two identical laser sources and FPGA breadboards were employed to simulate the transponder laser interferometric link between two satellites. Laser sideband communication was implemented using electro-optic modulators (EOMs), replicating on-orbit conditions. The test results demonstrated successful phase-locking of the two lasers, with no significant impact from laser sideband modulation. These findings provide essential technical validation for the development and deployment of future space-based gravitational wave detection.

\section*{Data availability statement}
The data cannot be made publicly available upon publication because no suitable repository
exists for hosting data in this field of study. The data that support the findings of this study are
available upon reasonable request from the authors.

\section*{Acknowledgments}
We thank members of Prof. Haixing Miao for useful discussions and the experimental facility support from the PGMF (National Precision Gravity Measurement Facility). This work was supported by the National Natural Science Foundation of China (Grant Nos. 12105375) and the National Key Research and Development Program of China (Grant Nos. 2022YFC2203901).

\section*{Reference}
\bibliography{ref}

\end{document}